# Study of spin dynamics and damping on the magnetic nanowire arrays with various nanowire widths

Jaehun Cho [a], Yuya Fujii [b], Katsunori Konioshi [b], Jungbum Yoon [c], Nam-Hui Kim [a], Jinyong Jung [a], Shinji Miwa [b], Myung-Hwa Jung [d], Yoshishige Suzuki [b], and Chun-Yeol You [a,*]

[a] Department of Physics, Inha University, Incheon, 402-751, South Korea
[b] Graduate School of Engineering Science,
Osaka University, Toyonaka, Osaka 560-8531, Japan
[c] Department of Electrical and Computer Engineering,
National University of Singapore, Singapore 117576
[d] Department of Physics, Sogang University, Seoul, 121-742, South Korea

**Abstract**

We investigate the spin dynamics including Gilbert damping in the ferromagnetic nanowire arrays. We have measured the ferromagnetic resonance of ferromagnetic nanowire arrays using vector-network analyzer ferromagnetic resonance (VNA-FMR) and analyzed the results with the micromagnetic simulations. We find excellent agreement between the experimental VNA-FMR spectra and micromagnetic simulations result for various applied magnetic fields. We find that the demagnetization factor for longitudinal conditions, $N_z$ ($N_y$) increases (decreases) as decreasing the nanowire width in the micromagnetic simulations. For the transverse magnetic field, $N_z$ ($N_y$) increases (decreases) as increasing the nanowire width. We also find that the Gilbert damping constant increases from 0.018 to 0.051 as the increasing nanowire width for the transverse case, while it is almost constant as 0.021 for the longitudinal case.

\* Corresponding author. FAX: +82 32 872 7562.

*E-mail address:* cyyou@inha.ac.kr

Ferromagnetic nanostructures have recently attracted much interest for the wide potential applications in high density spintronic information storage, logic devices and various spin orbit torque phenomena.[1,2,3,4,5] It is well known that the detail spin dynamics of nanostructure is far from the one of the bulk's because of many reasons, different boundary conditions, changes of the magnetic properties including the saturation magnetization, anisotropy energy, and exchange stiffness constant, etc. Since the magnetic properties are usually sensitive functions of the sample fabrication conditions, it has been widely accepted that the detail sample fabrications are also important in the study of spin dynamics. However, the relatively less caution has been made for the boundary conditions of the spin dynamics in the nanostructure.

In the spin transfer torque magnetic random access memory (STT-MRAM), the magnetic damping constant is important because the switching current density is proportional to the damping constant.[6] In the nanowire, damping constant also plays crucial role in the spin dynamics including domain wall motion with magnetic field[7] and spin transfer torque.[8] Furthermore, it is the most important material parameter in spin wave (SW) dynamics.[9] Despite of the importance of the damping constant, many studies about spin dynamics in ferromagnetic nanowires have not taken into account the damping constant properly.[10,11,12] Only a few studies paid attention to the magnetic damping in the nanowires spin

dynamics.[13,14]

In this study, arrays of CoFeB nanowires are prepared by e-beam lithography, and they are covered coplanar wave-guide for the ferromagnetic resonance (FMR) measurement as shown in Fig. 1. We measured FMR signal with longitudinal (wire direction) and transverse magnetic fields in order to investigate the spin dynamics with different boundary conditions. Also we extract Gilbert damping constant using micromagnetic simulations with the different applied magnetic field directions in various nanowire arrays. We find the damping constant decreases with increasing the nanowire width for the transverse magnetic field with constant input damping constant in micromagnetic simulations, while we obtain almost constant damping constant for the longitudinal field.

The films were prepared using DC magnetron sputtering. The stacks consist of Ta (5 nm)/$Co_{16}Fe_{64}B_{20}$ (30 nm)/Ta (5 nm) on single crystal MgO (001) substrates. The films are patterned as 100-nm-width wire arrays with 200-nm-space each wires using e-beam lithography and an Ar ion milling technique as shown in Fig. 1. The width is determined with a scanning electron microscope (SEM). These nanowire arrays are covered by coplanar wave guide in order to characterized with the Vector Network Analyzer (VNA)-FMR technique described elsewhere.[15] We prepare nanowire arrays as shown in Fig. 1, and external DC magnetic field direction for FMR measurement is also depicted.

We use VNA-FMR spectra to measure imaginary parts of the susceptibility of the samples.[16]

The measured imaginary parts of the susceptibility raw data are calibrated with the careful calibration procedures.[16] The calibrated imaginary parts of the susceptibility are shown in Fig 2(a) and (b) for an applied magnetic field at 0.194 T for the nanowire arrays. The un-patterned thin film is also examined for the reference. We find two resonance frequencies, 17.2 and 26.4 GHz, as shown in Fig. 2(a) for the nanowire array, while there is only one peak at 16.8 GHz for the un-patterned thin film as shown in Fig 2(b). We believe that the smaller peak (17.2 GHz) in Fig. 2(a) is originated from the un-patterned part of the nanowire array, because the frequency is closed to the un-patterned thin film's peak (16.8 GHz). Probably, the un-patterned part of the nanowire is formed due to poor e-beam lithography processes. On the other hand, the resonance frequency near 26.0 GHz is calculated from micromagnetic simulation at an applied magnetic field at 0.200 T, as shown in Fig. 2 (c). We clarify the source of the main peak (26.4 GHz) is nanowire arrays by using micromagnetic simulation. These two peaks named as the uniform FMR mode (smaller peak position) and nanowire mode (higher peak position).

In order to determine the saturation magnetization, the resonance frequencies are measured as a function of the applied magnetic field, and the results are fitted with the Kittel's equation.[17] This equation employs the corresponding demagnetization factors of $N_x = 0$, $N_y = 0$ and $N_z = 1$ for un-patterned film, when applied magnetic field $H$ is $x$- direction with following equations,

$$f = \frac{\gamma}{2\pi}\sqrt{\{H+(N_y-N_x)M_s\}\{H+(N_z-N_x)M_s\}}. \quad (1)$$

Here, $\gamma$ is the gyromagnetic ratio, $H$ is the applied magnetic field, $M_s$ is saturated magnetization, $N_x$, $N_y$, and $N_z$ are the demagnetization factors applying the cyclic permutation for the applied magnetic field direction.

The micromagnetic simulations are performed by using the Objective-Oriented-MicroMagnetic Framework (OOMMF)[18] with 2-dimensional periodic boundary condition (PBC).[19] We select a square slat of 100 nm × 100 nm × 30 nm nanowire separated 200 nm in $y$- direction with a cell size of 5 nm × 5 nm × 30 nm. The material parameters of CoFeB used in our simulation are summarized as follows: $M_s$ = 15.79 × 10$^5$ A/m, the exchange stiffness 1.5 × 10$^{11}$ J/m, the gyromagnetic ratio 2.32 × 10$^{11}$ m/(A·s) and we ignore the magneto-crystalline anisotropy. In this simulation, the Gilbert damping constant of 0.027 is fixed. The saturation magnetization and Gilbert damping constant are determined by using VNA-FMR measurement for un-patterned thin film. For the exchange stiffness constant, experimentally determined values are range of 0.98 to 2.84 × 10$^{11}$ J/m which value has dependence on the fabrication processes[20] and composition of ferromagnetic materials,[21] while we have picked 1.5 × 10$^{11}$ J/m as the exchange stiffness constant. The determination method of Gilbert damping constant will be described later.

In order to mimic FMR experiments in the micromagnetic simulations, a "sinc" function $H_y(t) = H_0 \sin\left[2\pi f_H (t-t_0)\right] / 2\pi f_H (t-t_0)$, with $H_0$ = 10 mT, and field frequency $f_H$ = 45 GHz, is applied the whole nanowire area.[22] We obtain the FMR spectra in the corresponding frequency range from 0 to 45 GHz. The FMR spectra due to the RF-magnetic field are obtained by the fast Fourier transform (FFT) of stored $M_{y(x)}$ (x, y, t) in longitudinal (transverse) $H_0$ field. More details can be described elsewhere.[23]

The closed blue circles in Fig. 3 is the calculated values with the fitting parameter using Eq. (1) which are fitted with the experimental data of un-patterned thin film. The obtained $M_s$ is 15.79 × 10$^5$ A/m while gyromagnetic ratio is fixed as 2.32 × 10$^{11}$ m/(A·s). The obtained $M_s$ value is similar with vibrating sample magnetometer method[24] which CoFeB structure has Ta buffer layer. The resonance frequencies of uniform FMR mode in nanowire arrays are plotted as open red circles in Fig. 3. The resonance frequencies of uniform FMR mode is measured by VNA-FMR are agreed well with resonance frequency of un-patterned thin film measured by VNA-FMR. In Fig. 3, the applied field dependences of the resonance frequencies Measured by VNA-FAM for the nanowire are plotted as open black rectangular, along with the result of micromagnetics calculated with Eq. (1) as depicted closed black rectangular. It is also well agreed with the experimental result in nanowire mode and micromagnetic simulation result in the nanowire arrays.

In order to reveal the effects of spin dynamics properties with various nanowire widths, we perform micromagnetic simulations. The nanowire widths are varied from 50 to 150 nm in 25-nm step for fixed 200-nm-space with PBC, it causes changes of the demagnetization factor of the nanowire. In Fig. 4 (a) shows the nanowire width dependences of the resonance frequencies for the longitudinal magnetic field (open symbols) along with the resonance frequencies calculated with Eq. (1) (solid lines). The demagnetization factors can be determined by fitting Eq. (1) while $N_x$ is fixed as 0 to represent infinitely long wire. The agreements between the results of micromagnetic simulations (open circles) and Eq. (1) (solid lines) are excellent.

For the transverse magnetic field, the direction of applied magnetic field is $y$ - axis, Eq. (1) can be rewritten as follows:

$$f = \frac{\gamma}{2\pi}\sqrt{\left(H+\left(N_x-N_y\right)M_s\right)\left(H+\left(N_z-N_y\right)M_s\right)} \quad . \tag{2}$$

In this equation, we use the relation of demagnetization factors, $N_x + N_y + N_z = 1$, in order to remove uncertainty in the fitting procedure. In the transverse field, the demagnetization factors are determined by Eq. (2). The resonance frequencies for transverse magnetic field which are obtained by micromagnetic simulation (open circles) and calculated by Eq. (2) (solid lines) as a function of the applied magnetic field with various nanowire width are displayed in Fig. 4(b). The longitudinal case, when the field direction is easy axis, they are saturated with small field. However, the transverse case, when the field direction is hard axis, certain amount of field is necessary to saturate along the transverse directions. The narrower nanowire, the larger field is required as shown in Fig. 4 (b).

Fig. 5(a) and (b) show the changes of demagnetization factors in longitudinal and transverse magnetic fields as a function of the nanowire width, respectively. The demagnetization factors play important role in the domain wall dynamics, for example the Walker breakdown is determined by the demagnetization factors.[25] Furthermore, they are essential physical quantities to analyze the details of the spin dynamics. It is clearly shown that the $N_z$ ($N_y$) increases (decreases) with increasing the nanowire width in longitudinal magnetic field. For the transverse magnetic field, $N_z$ ($N_y$) increases (decreases) with increasing the nanowire width, during $N_x$ is almost zero value. The demagnetization factors both longitudinal and transverse have similar tendency with the effective demagnetization factors of dynamic origin[26] and the static demagnetization factors for the prism geometry.[27]

Now, let us discuss about the Gilbert damping constant $\alpha$. The relation of the full width and half maxima ($\Delta f$) of a resonance peaks as a function of applied field are shown in Fig. 6 for longitudinal (a) and transverse (b). The $\Delta f$ is given by[15]:

$$\Delta f = \alpha \frac{\gamma}{2\pi} \left( 2H + M_s \left\{ \left( N_{y,(x)} + N_z \right) - \frac{N_{x,(y)}}{2} \right\} \right) + \Delta f_{ex} . \quad (3)$$

where, $\Delta f_{ex}$ is the extrinsic line width contributions, when the applied magnetic field is *x*-(*y*-)axis for longitudinal (transverse) case. The symbols are the results of the micromagnetic simulations and the solid lines are the fitting result of Eq. (3). We use pre-determined demagnetization factors (Fig. 5) during fitting procedures, and the agreements are excellent.

We have plotted the Gilbert damping constant as a function of the wavevector in nanowire width ($q = \pi / a$, *a* is the nanowire width) in Fig. 7. The black open rectangles are data extracted from the transverse field and the red open circles are longitudinal field data. We find that the Gilbert damping constant varied from 0.051 to 0.018 by changing the wavevector in nanowire width in transverse field. On the other hand, longitudinal field case the damping constant is almost constant as 0.021. Let us discuss about the un-expected behavior of the damping constant of transverse case. The wire width acts as a kind of cut-off wavelength of the SW excitations in the confined geometry. SWs whose wavelength are larger than 2*a* are not allowed in the nanowire. Therefore, only limited SW can be excited for the narrower wire, while more SW can be existed in the wider wire. For example, we show transverse standing SW as profiled in the inset of Fig. 6 for 150-nm width nanowire in our micromagnetic simulations. More possible SW excitations imply more energy dissipation paths, it causes larger damping constant. For narrower nanowire (50-nm), only limited SWs can be excited, so that the damping constant is smaller. However, for the limit case of infinite *a* case, it is the same with un-patterned thin films, there is no boundary so that only uniform mode can be excited, the obtained damping constant must be the input value.

In summary, the VNA-FMR experiments is employed to investigate the magnetic properties of CoFeB nanowire arrays and the micromagnetic simulations is proposed to understand the magnetic properties including Gilbert damping constant of various CoFeB nanowire arrays width. We find that the demagnetization factors are similar with the dynamic origin and static for the prism geometry. The wire width or SW wavevector dependent damping constants can be explained with number of SW excitation modes.

**ACKNOWLEDGMENTS**

This work was supported by the National Research Foundation of Korea (NRF) Grants (Nos. 616-2011-C00017 and 2013R1A12011936).

# Figure Captions

Fig. 1. Measurement geometry with SEM images of the 100-nm-width nanowires with a gap of 200 nm between nanowires. The longitudinal nanowire arrays are shown. After the nanowire patterns have been defined by e-beam lithography, they are covered by co-planar wave guides.

Fig. 2. (a) The measured FMR spectrum of the CoFeB nanowire with H =0.194 T. The red (lower peak) and blue (higher peak) arrows indicate the resonance frequencies of the uniform FMR mode and the nanowire mode, respectively. (b) The measured FMR spectrum of the CoFeB thin film with H =0.194 T. (c) Simulated FMR spectrum of the CoFeB nanowire with H= 0.200 T.

Fig. 3. Measured and calculated FMR frequencies with the applied magnetic field for 100-nm-width nanowire. The open black rectangles are nanowire mode and open red circles are the uniform FMR mode for CoFeB thin film. The closed black rectangles are calculated by OOMMF and the closed blue circles are theoretically calculated by Eq. (1) using fitted parameters form un-patterned film.

Fig. 4. Variation of resonance frequencies with the applied magnetic field for the different PBC wire width for (a) longitudinal field and (b) transverse field. Inset: The geometry of 2-dimensional PBC micromagnetic simulation with nanowire width a and a gap of 200 nm between nanowires. The black open rectangles, red open circles, green open upper triangles, blue open down triangles, cyan open diamonds represent as nanowire width as 50 nm, 75nm, 100 nm, 125nm, and 150 nm, respectively.

Fig. 5. Demagnetization factor with PBC wire width for (a) longitudinal and (b) transverse field. The black open circles, red open rectangles, blue open upper triangles represent as demagnetization factors, $N_y$, $N_z$, and $N_x$, respectively.

Fig. 6. Full width and half maxima with the applied magnetic field for (a) longitudinal and (b) transverse field. The black open rectangles, red open circles, green open upper triangles, blue open down triangles, cyan open diamonds represent as nanowire width as 50 nm, 75nm, 100 nm, 125nm, and 150 nm, respectively.

Fig. 7. Damping constants with wavevector for transverse (the black open rectangles) and longitudinal (the red open circles) field with errors. The black line is the input value which is determined from un-patterned film. Inset presents the profile of the transverse spin density as SWs.

Fig. 1

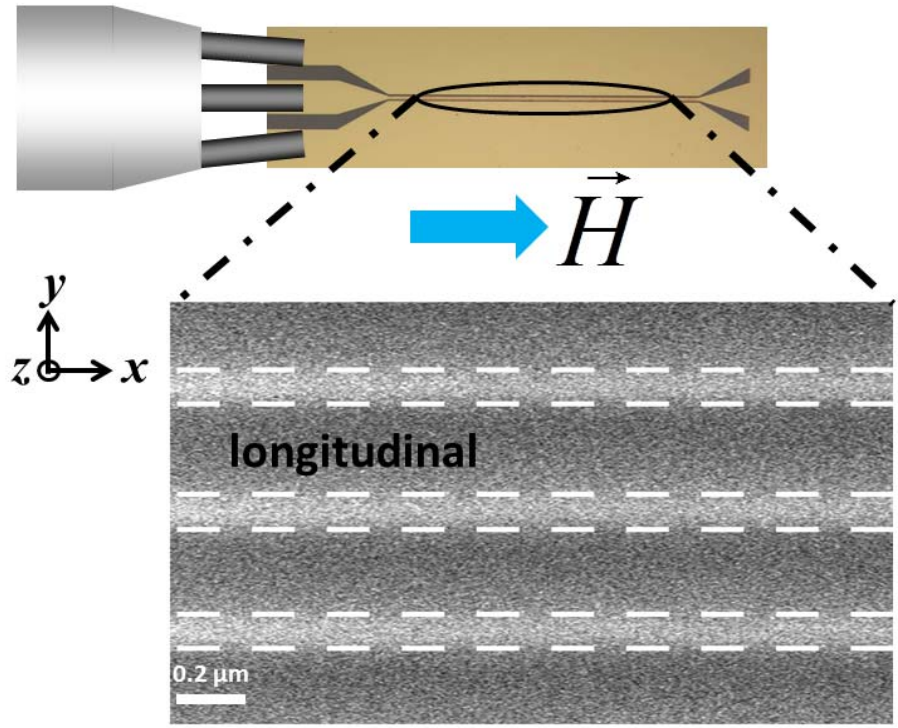

Fig. 2.

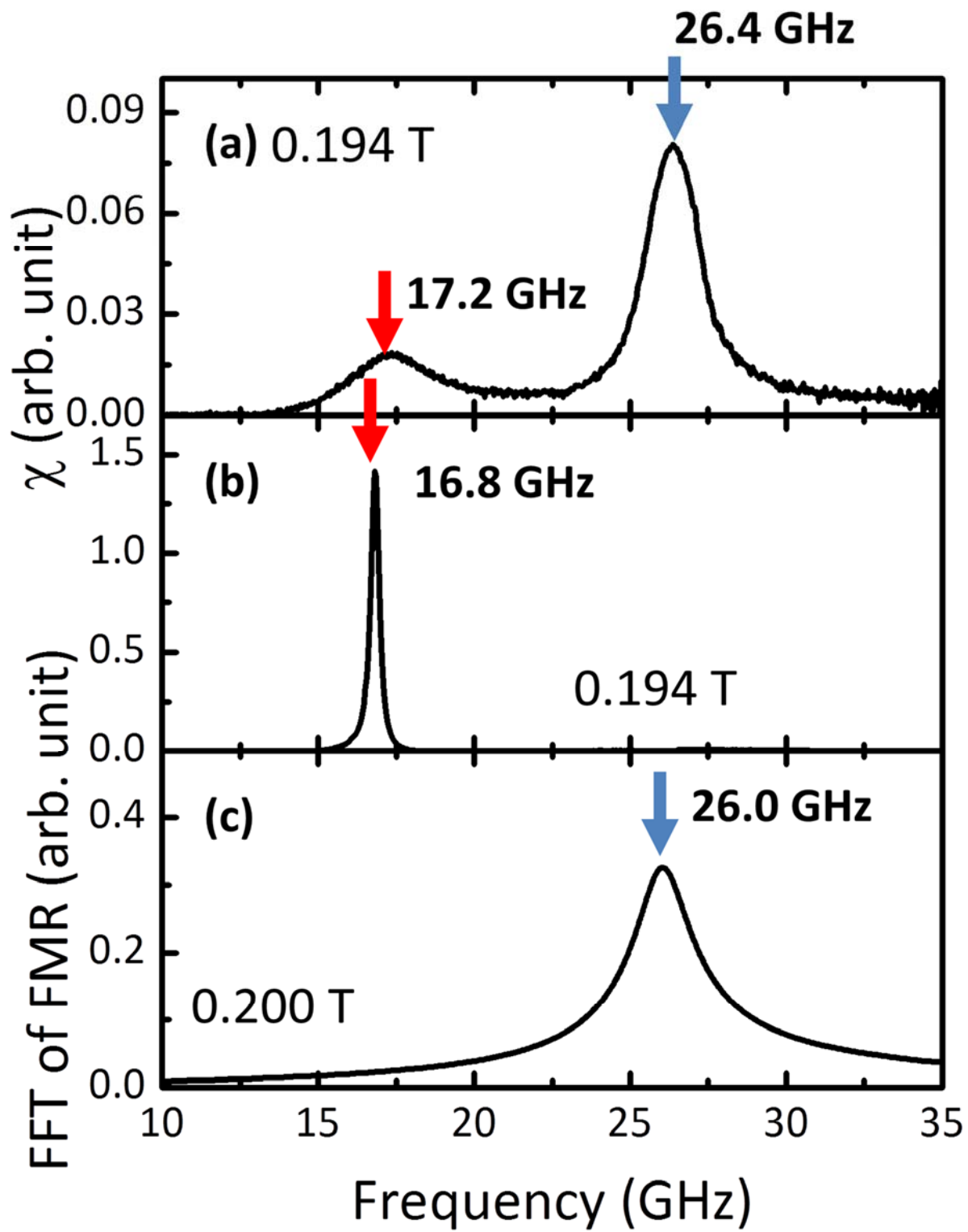

Fig. 3.

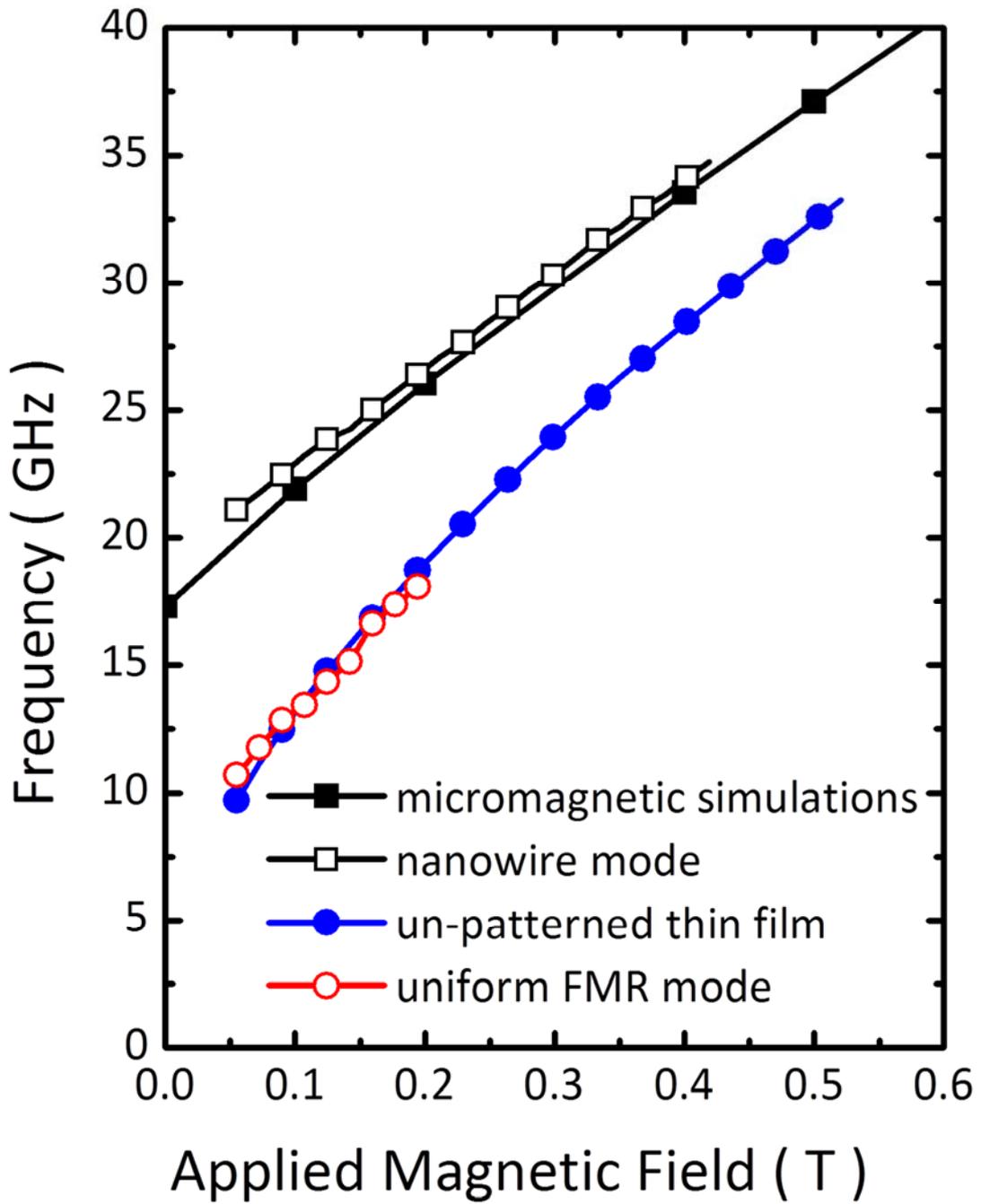

Fig. 4

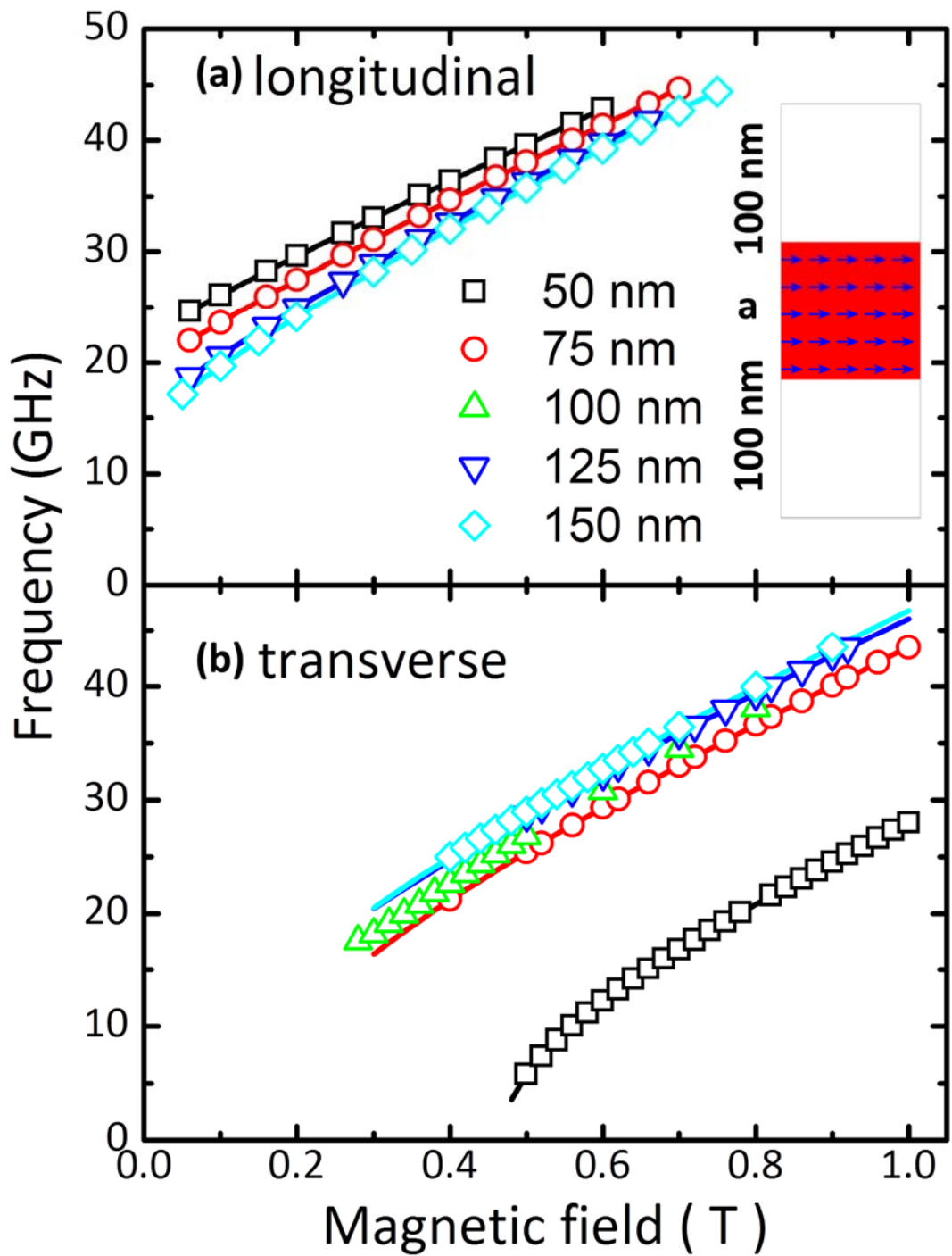

Fig. 5

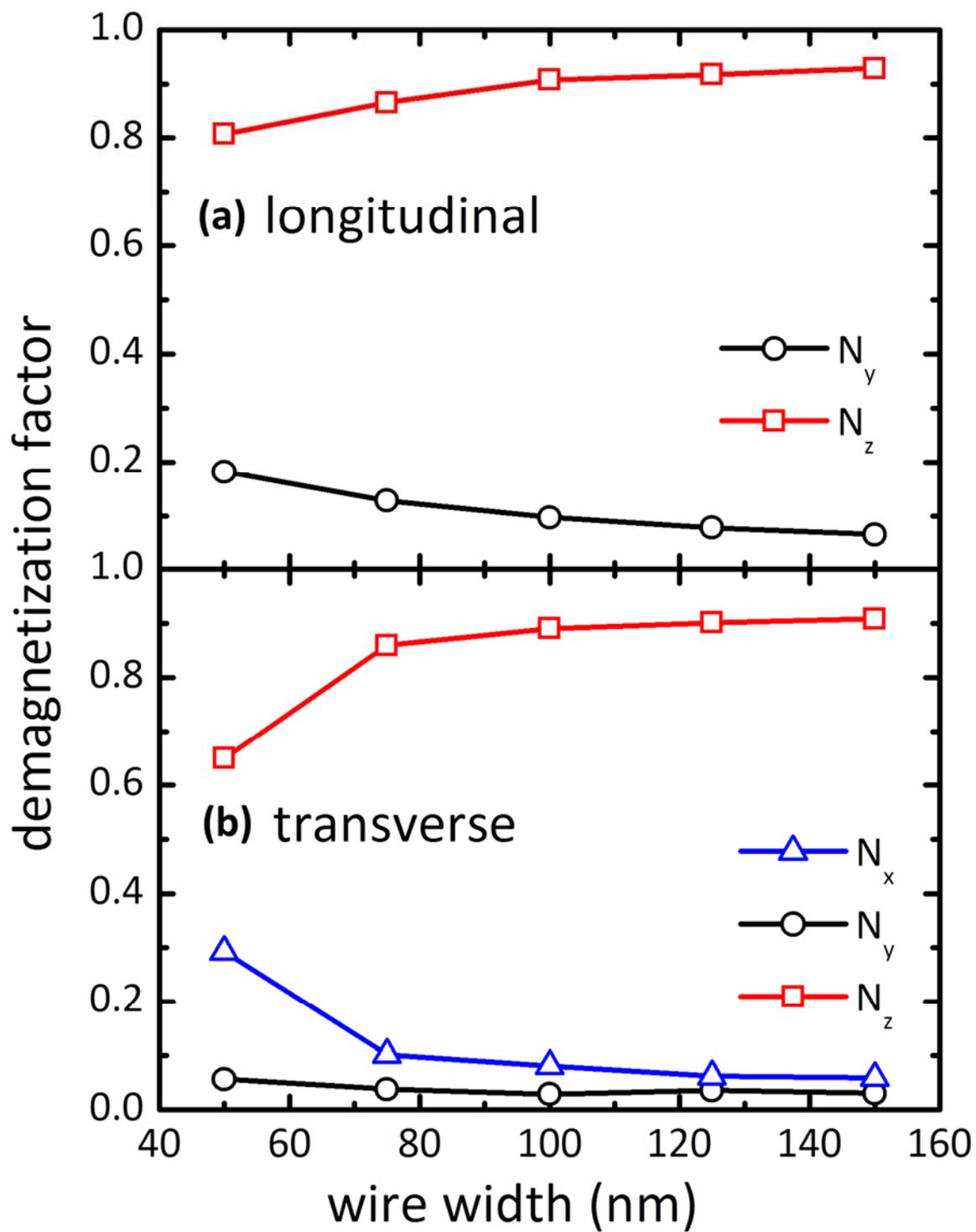

Fig. 6

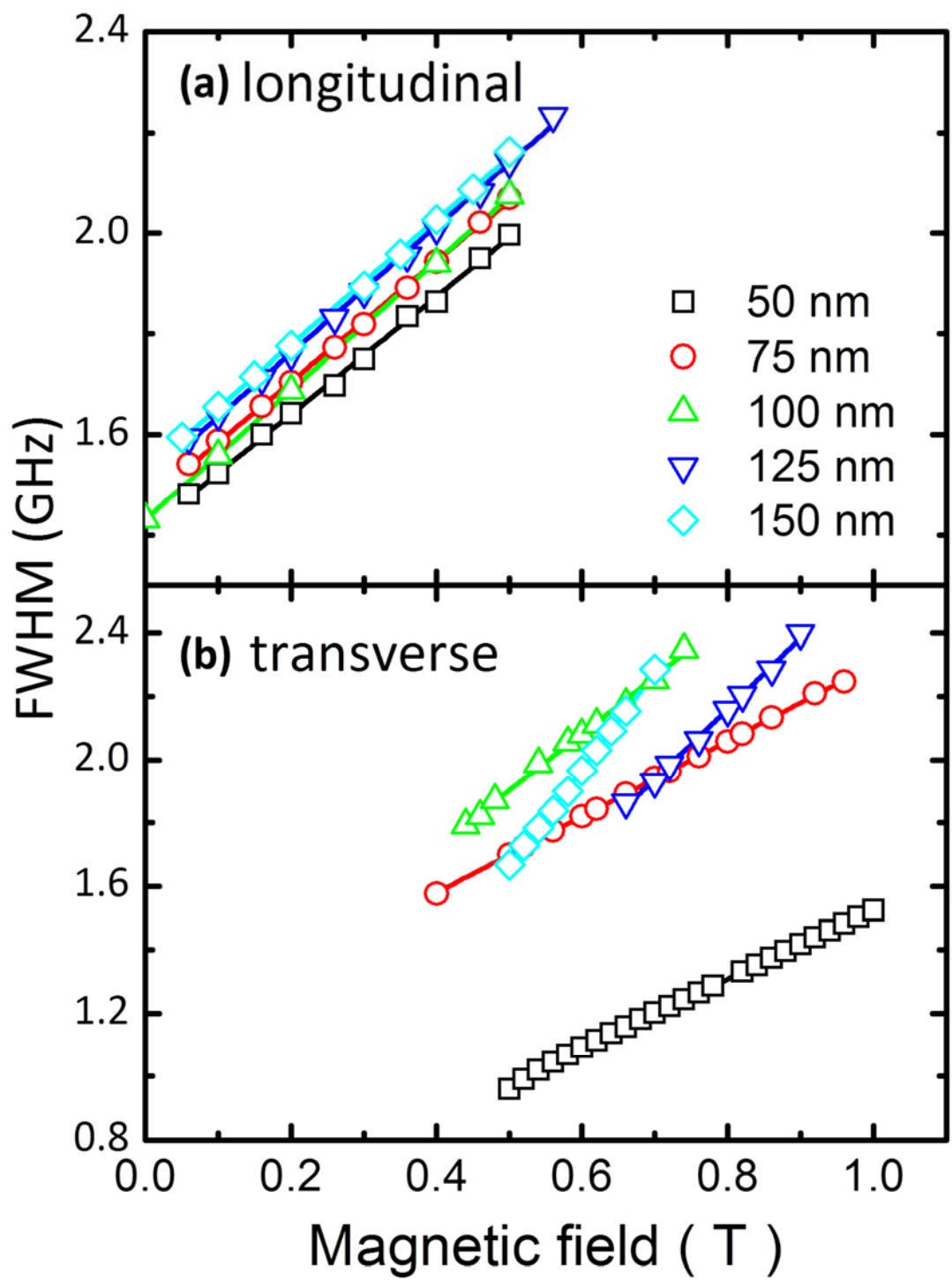

Fig. 7

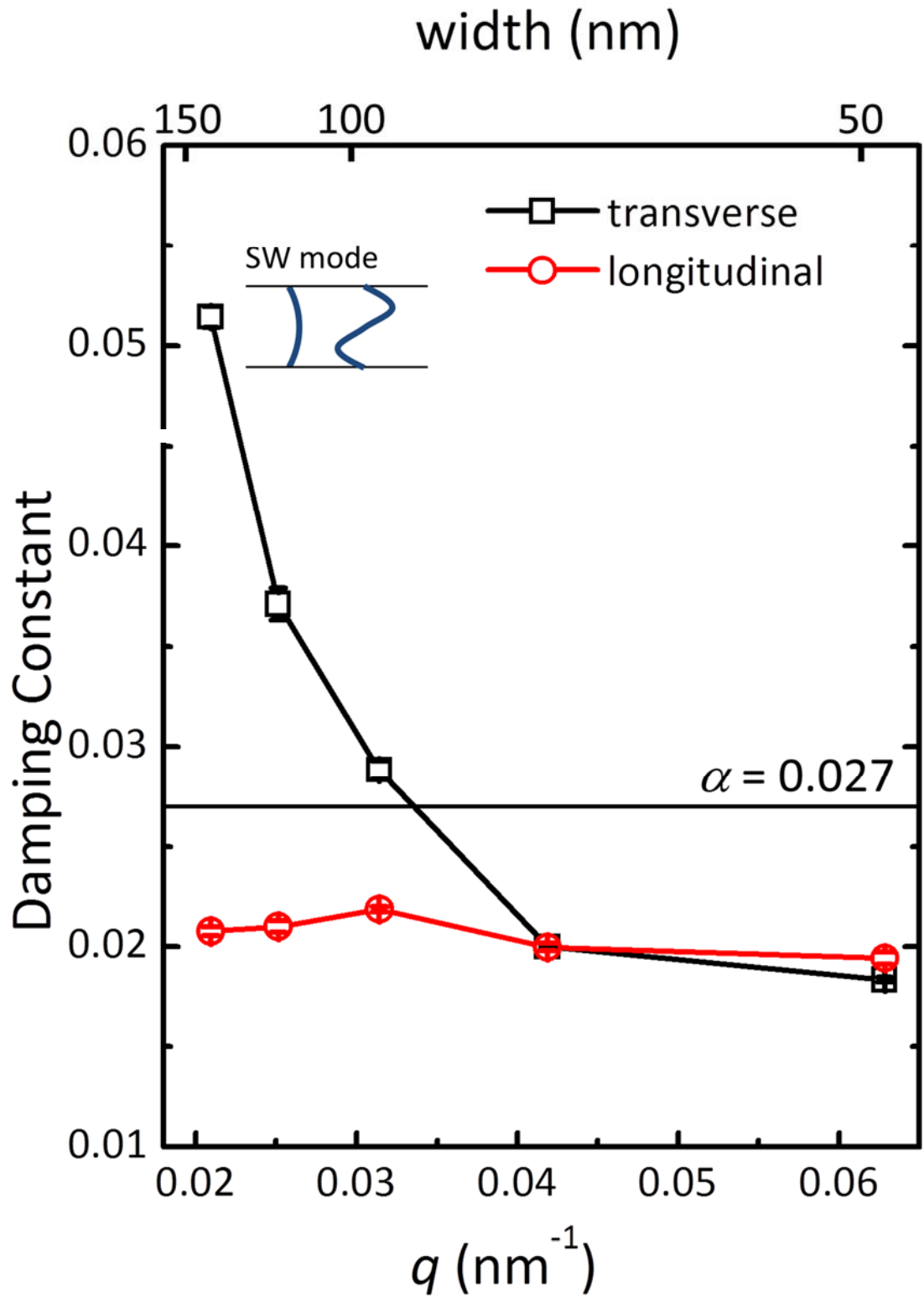